\documentclass[conference]{IEEEtran}

\usepackage[colorlinks,bookmarksopen,bookmarksnumbered,linkcolor=black,citecolor=black,urlcolor=black]{hyperref}
\usepackage{url}
\usepackage{tabularx}
\usepackage{multirow}
\usepackage{cite}
\usepackage{rotating}
\usepackage{color}
\usepackage[latin1]{inputenc}
\usepackage{graphicx,subfigure}%

\usepackage{amsmath}
\usepackage[linesnumbered,ruled]{algorithm2e}

\begin{document}

\title{Hybrid Wired-Wireless Backhaul Solutions for Heterogeneous Ultra-Dense Networks}
\author{
	\IEEEauthorblockN{Onel L. A. L\'opez, Hirley Alves, Richard Demo Souza\IEEEauthorrefmark{1}, Matti Latva-aho}
	\IEEEauthorblockA{Centre for Wireless 	Communications (CWC), Oulu, Finland\\}
	\IEEEauthorblockA{\IEEEauthorrefmark{1}Federal University of Santa Catarina (UFSC), Florianopolis, Brazil\\}
				firstname.lastname@oulu.fi, \IEEEauthorrefmark{1}richard.demo@ufsc.br
}

\maketitle
\begin{abstract}
Wireless networks are becoming extremely pervasive while traffic demand is ever increasing. In order to cope with the forecast increase in traffic volume for the upcoming years, as well as the number of connected devices, new technologies, practices and spectrum rearrangements are required.  
In this context, a key question arises: {how to provide extensive {backhaul} connectivity and capacity for pervasive ultra dense networks?}
The answer is rather complex, if feasible. To shed some light into this issue we overview potential technologies, either wired or wireless, and identify  technical challenges. 
Moreover, we evaluate an illustrative scenario of a ultra-dense network that operates with hybrid wired-wireless backhaul. We assume multiple radio access technologies at small and macro base stations (BSs), and we discuss optimal traffic splitting and routing solutions for different topologies and traffic profiles. 
\end{abstract}
\section{Introduction}
Wireless networks have become ubiquitous and an indispensable part of our daily life. There is an ever increasing demand for wireless data services; a thousand-fold increase in traffic volume is expected by 2020. In addition, billions of smart devices will be connected to the Internet, with potential to generate new business models and an economic impact in the order of trillions of dollars. To cope with these demands, wireless industry is searching for new ways to improve coverage and capacity, while lowering their capital and operating expenditures \cite{Ericsson2015, Boccardi5GCM2014}. 

Despite the recent advances, current wireless technologies cannot withstand such large increase in traffic and user density, and hence new practices and spectrum rearrangements are required. Wireless networks architecture are evolving to become more scalable, flexible, heterogeneous and dynamic in order to offer tailored and optimized solutions \cite{Ericsson2015,Boccardi5GCM2014}. Current networks are already heterogeneous to some extent, and that will be inherent of future generations \cite{Boccardi5GCM2014}.  
A heterogeneous ultra dense network (UDN) is composed of several low cost and compact small cell access points, in conjunction with  macro cell BSs. The small cell access points are low power BSs, and hence have limited coverage compared to macro BSs, but can be easily deployed. The UDN paradigm presents advantages of easy installation and reduced cost, which motivates a dense deployment, though techniques to handle interference and aggressive frequency reuse are still needed \cite{QuekTWC2015}. Another feature of the current and future wireless networks is the coexistence of several radio access technologies (RATs) within small and macro BSs \cite{Ericsson2015}. Some early works evaluate UDNs with co-located LTE and WiFi RATs \cite{GalininaJSAC2015} evincing the throughput gains of splitting traffic into several RATs. 
UDNs and multiple RATs deployments are paradigm shifts for wireless networks, and therefore there are many open related research challenges \cite{Ericsson2015}, such as interference management, decentralized RAT and path selection, prioritization of traffic flows given the QoS and the inherent constraints of each air interface, such as delay profile and achievable rates. 

The backhaul connects the small and macro BSs to the core network~\cite{ART:Jaber-ACCESS16}. Macro cell backhaul is often connected via fiber, due to its large capacity and bandwidth, low latency, but at a cost of elevated operating expenditures. UDNs pose challenges with respect to positioning, since the small cells deployment may be indoors or outdoors, at the street level, even in locations hard to be reached by dedicated fiber or other copper wired alternatives. 
In this context, wireless backhaul appears as a way to complement the wired solutions, due to easy installation, availability, and planning, which reduce deployment costs. Due to spectrum scarcity, new frequency bands, especially millimeter wave (mmWave) \cite{Ericsson2015, Boccardi5GCM2014}, appear as viable solutions for both access and dedicated wireless backhaul links, due to large bandwidths and data rates attained. 

Backhaul is a key enabler of future wireless networks, such as 5G, but if not addressed properly it becomes its Achilles heel \cite{ieeeCTN2015}. Therefore, a key question arises: {how to provide extensive backhaul capacity and connectivity in UDNs?} This is a complex and open problem to current deployments and becomes even more intricate in UDNs with multiple RATs.  

The key contributions of this paper are twofold: \textit{i)} to (briefly) overview current backhaul solutions; \textit{ii)} to evaluate the convenience of hybrid backhaul in UDNs, in which each node has multiple RATs, wired or wireless. We assess the average delay and throughput of an illustrative network topology subject to throughput and average delay of each interface. The proposed algorithm splits the incumbent traffic from a new user among the different interfaces guaranteeing quality of service requirements. Our results demonstrate the feasibility of a hybrid solution exploiting wireless and wired technologies in dense networks, even in the advent of link failures. 

Next, in Section~\ref{sc:back-net} we overview the envisaged backhaul for an UDN deployment and assess the most promising wired and wireless technologies. Then, in Section~\ref{st:Hybrid} we introduce our solution and some illustrative numerical results. Finally, Section~\ref{sect:remarks} concludes the paper. 
\section{Backhaul Networks}\label{sc:back-net}
Assume that both macro and small cells have one to several RATs available, both wireless and wired connections, and we consider that traffic is split among the available interfaces. The UDN snapshot shows possible connections between small cells, the macro cell and the core network.
The concept of a centralized radio access network (C-RAN) has emerged recently. In this scenario the \textit{backhaul} becomes a complex network split into: fronthaul, midhaul and backhaul.
The C-RAN objective is to achieve very low latency between BSs, known as remote radio heads (RRH), by splitting their functions and allocating them to a centralized pool of baseband resources, the baseband unit (BBU). In this context, the fronthaul connects the RRH to the BBU, or to an aggregation point, while the midhaul, whenever needed, connects the fronthaul aggregation point to the BBU. By its turn, the backhaul connects BBUs to the core network \cite{ART:Jaber-ACCESS16}. Distinct requirements exist for front, mid, and backhaul, while a wide range of candidate solutions that comply to some of these requirements are available. 
The backhaul network, including front, mid and backhaul, is currently covered by microwave, fiber and copper-wire links, though those options are not able to cope with the future exponential increase in volume of data traffic, number of connected devices, demand for ultra-high quality video transmissions and low latency for real-time applications such as gaming and augmented reality \cite{ART:Allen-Vodafone14a, ART:SCF-2013}. These diverse requirements impact on the design of the backhaul network, which needs to be flexible and dynamic to efficiently comply with stringent and time varying demands, and evinces the need for complementary wired-wireless technologies leading to hybrid backhaul networks.
Next, we overview the most promising solutions for hybrid backhauling.
\vspace{-2mm}
\subsection{Wired Candidates for UDN backhaul}\label{sect:wired}

One of the most suitable technologies for backhauling is point-to-point fiber, which offers Gbps throughput and low latency, fitting into many applications and complying to many of the requirements of future networks. However, it comes with high leasing fees. Alternatively, Fiber To The X (FTTX) encompasses a family of fiber optic access architectures, whose capacity and coverage changes for different variants, but overall hundreds of Mbps up to some Gpbs are attainable with lower costs than direct fiber \cite{ART:SCF-2013}, but both suffer from lack of sufficient installed infrastructure.
Due to the increased density in UDN, commonly used personal broadband solutions, such as  Digital Subscriber Line (xDSL), appear as an third option and an attractive solution because of its large installed infrastructure, relative low cost of installation, operation and maintenance compared to fiber solutions \cite{ART:SCF-2013}. 
On the other hand, xDSL falls short in terms of throughput (up to hundreds of Mbps) compared to optical solutions, though range may reach a few kilometers \cite{ART:Jaber-ACCESS16}. Some variants of xDSL family offer higher throughput, but with limited coverage.

Depending on the application, a hybrid copper-fiber solution may be a cost-effective option. 
Nonetheless, as highlighted in \cite{ART:Jaber-ACCESS16}, fiber solutions are the most attractive but installation is expensive and may discourage operators. However, when associated with wireless solutions a good compromise in terms of coverage, latency and data rates can be attainable.
\vspace{-2mm}
\subsection{Wireless solutions for UDNs}\label{sect:wireless}
Currently, Microwave links are used as backhaul solutions and operate mostly in the range between $10$ to $28$ GHz. Those links present a good trade-off between capacity and available spectrum, but operate under line-of-sight (LOS) conditions through directional antennas with fixed alignments \cite{ART:CBNL-2014}. Microwave solutions comply with ultra reliable low latency communications, with a $1$ millisecond round trip latency per hop and coverage of few kilometers.

Alternatively, mmWave solutions can potentially cope with harsh demands of Gbps throughput and extremely low latency, which has motivated researchers from industry and academia to use mmWave spectrum for broadband communications, due to larger amounts of available bandwidth and therefore the potential to achieve Giga-bit throughput and millisecond latency. For instance, around $60$ GHz, up to $9$ GHz of bandwidth are available~\cite{ART:SCF-2013}. Recent results propose new array architectures and show that enhanced performance and coverage is attained via beamforming, including beamsteering and beam alignment, at both transmitter and receiver \cite{ART:Pi-CMag2016}. In general, weather impairments can cause severe attenuation in mmWave bands, impacting coverage, which is quite limited even under LOS conditions, reaching up to a few hundred meters. Solutions based on mmWave are not only seen as a key technology for backhauling, but also for broadband applications where extremely high data rates are needed to cope with the crescent traffic demand \cite{Ericsson2015}. 

Most of current wireless networks also rely on sub-$6$~GHz spectrum, whose key characteristic is good propagation even in nLOS, which is one of the drawbacks of mmWave and Microwave counterparts despite the recent advances. However, spectrum fragmentation, in terms of licensing, is an issue, since regulations vary from country to country, as well as the requirement of sophisticated interference cancellation and avoidance mechanisms, besides the high licensing costs. Thus, sub-$6$~GHz appears as a complementary solution to mmWave and Microwave due to the available installed infrastructure. 

One other option, is Satellite communication, which is advantageous in terms of availability and worldwide coverage. On the other hand, the major drawback is its extremely high latency, in the order of hundreds of milliseconds, associated with low throughput. Besides, the costs are calculated based on the consumed Mbps. Satellite communication becomes an option for locations where the cost of laying a wired solution, or having multiple wireless hops, overcomes the cost of the consumed Mbps. 

\vspace{-2mm}
\section{Hybrid multiple RATs backhaul for UDNs}\label{st:Hybrid}
\begin{figure}[!t]
	\centering
	\includegraphics[width=.75\columnwidth]{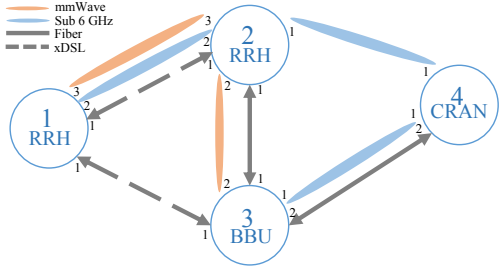}
	\vspace{-5mm}
	\caption{Network deployment composed of two RRHs, BBU and CRAN. We assume at least one wired and/or wireless RATs are available at each node, as well as mean delay and throughput as indicated on the right-hand side.}
	\label{fig:multi_rat}
	\vspace{-6mm}
\end{figure}

Consider a network topology where nodes can be interconnected through several link technologies, wired and/or wireless. Fig.~\ref{fig:multi_rat} exemplifies a possible topology in ultra dense networks and will be used to exemplify the robustness of the algorithm described next. Fig.~\ref{fig:multi_rat} shows a network deployment composed of two RRHs, BBU and CRAN each with one or multiple RATs. We assume at least one wired and/or wireless RATs are available at each node.
Let $ijk$ be the link identifier of the connection between nodes $i$ and $j$ through interface $k$, which is characterized through the link capacity $c_{ijk}$, mean occupation $o_{ijk}$, and mean delay $d_{ijk}$. A new user, which is served by one of the access nodes, e.g., a RRH or small cell BS, requires a connection with the following QoS requirements: transmission rate $r_0$ and maximum allowable delay $d_0$. Then, a question arises: {What is the recommended path for the new user traffic given its QoS requirements, $r_0$ and $d_0$, and the network traffic profile, $\{c_{ijk},o_{ijk},d_{ijk}\}$?}

\begin{algorithm}\label{alg_1}
	\SetKwInOut{Input}{Input}
	\SetKwInOut{Output}{Output}
		
	\Input{
		\begin{itemize}
			\item Characteristics of the links: $o_{ijk}$, $d_{ijk}$, $c_{ijk}$ $\forall i,j, k$.
			\item QoS requirements of the user to be allocated: $r_0$, $d_0$.
		\end{itemize}
		}
	\Output{
		\begin{itemize}
			\item Optimum path: $\{(n_i,n_{i+1},k^*_{n_in_{i+1}})\}$, $1\le i\le \nu$.
		\end{itemize}	
	\hrulefill
	} Form the multigraph $G = (V, W)$ with link weights following the next rule

	\eIf{$c_{ijk}(1-o_{ijk})>r_0$}
	{
		$w_{ijk}=o_{ijk}^md_{ijk}$
	}
	{
		$w_{ijk}=\infty$
	}
    Transform $G$ into a simple graph $G'=(V, W')$ with link weights $w_{ij}=\min_kw_{ijk}$, so $k^*_{ij}=\arg\min_kw_{ijk}$ is the interface selected for the $ij$ connection.
    
    Use Dijkstra's algorithm to find the path in $G'$ with minimum sum-weight. The optimum path is the collection of triplets $\{(n_i,n_{i+1},k^*_{n_in_{i+1}})\}$, $1\le i\le \nu$, where $n_i$ is the $i$th node in the optimum path and $\nu$ is the number of edges of it.
    
    \uIf{$\sum\limits_{i=1}^{\nu}d_{n_in_{i+1}k^*_{n_in_{i+1}}}\le d_0$}
    {
    	END: The path was found!
    }
    \uElseIf{$m>0$}
    {
    	Decrease $m$ ($m\ge 0$) and go back to Step 1.
    }
    \Else{Split the data into fragments with required transmission rate of $\{r_j\}$ such that $\sum_{j}r_j=r_0$ and run the Algorithm from step 1 for each fragment.}  
	\caption{Optimization algorithm}
\end{algorithm}
The backhaul network can be modeled through a weighted multigraph $G = (V, W)$, where $V$ is the set of vertexes and $W$ is the set of their weights according to some metric. Since we are interested in the uplink, there is no need to distinguish the link $ijk$ from $jik$, and the graph is not required to be directed. We use a Dijkstra-based algorithm, Algorithm~\ref{alg_1}, to compute the optimum path to reach the C-RAN (node 4), based on mean and not so-frequently updated statistics, which is a valid assumption if the traffic is correlated over time. In that sense, it seems convenient to avoid using heavy-loaded links to prevent routing failure due to path congestion. Aiming at that, the link weights are taken as $w_{ijk}=o_{ijk}^md_{ijk}$ if $c_{ijk}(1-o_{ijk})>r_0$ and $w_{ijk}=\infty$ otherwise, where $m\ge 0$ is a trade-off factor weighting the link rate availability. A larger $m$ accounts for a heavier impact on the link utilization, and $m = 0$ ignores it completely while computing the minimum delay path.
The Dijkstra algorithm can find the shortest path from a source node to a destination node in a simple graph, as well as the corresponding cost. However, this algorithm can be easily extended to the multigraph case by transforming the multigraph into a simple graph~\cite{Biswas.2013} $G'=(V, W')$ with weights chosen as $w_{ij}=\min_kw_{ijk}$, while $k^*_{ij}=\arg\min_kw_{ijk}$ is the interface selected for the $ij$ connection. Then, the optimum path between the serving node and the C-RAN can be computed through the Dijkstra algorithm. If the found path offers a delay greater than $d_0$, $m$ can be decreased and the algorithm executed again until a valid result is found. If a valid result is not found, the data can be split into fragments for which the algorithm can be executed independently. Notice that the steps of decreasing $m$ and splitting the data are not required to be necessarily in that order, however analyzing the best strategy is out of the scope of this work, and we just only follow the mentioned order.

\subsection{Numerical Example}\label{numerical_example}
Next, we provide a numerical example and therefore we resort to the network topology shown in Fig.~\ref{fig:multi_rat}. We assume two distinct QoS user profiles as follows: \textit{i)} $p_1=\{r_0=8\mathrm{Mbps},\ d_0=30\mathrm{ms}\}$ which includes a stringent latency requirement; and \textit{ii)} $p_2=\{r_0=30\mathrm{Mbps},\ d_0=50\mathrm{ms}\}$ where throughput is the key metric. We assume a minimum allowed rate of $0.2$ Mbps, which corresponds to the minimum packet fragmentation size. Our results come from averaging the algorithm execution outputs over $5000$ randomly generated traffic-load configurations. Based on the deployment shown in Fig.\ref{fig:multi_rat}, the mean delay and throughput used for numerical analysis are summarized in Table \ref{tb:delay-tput} \cite{ART:Jaber-ACCESS16}.
\begin{table}[!t]
	\caption{RATs throughput and mean delay}
	\label{tb:delay-tput}
	\centering
	\begin{tabular}{c|c|c|c}
		Technology & Links & Throughput (Gbps) & Mean delay (ms) \\
		\hline \hline
		Fiber & 231, 342 & 2 & 5 \\
		xDSL & 121, 131 & 0.05 & 20 \\
		Sub-6GHz & 122, 241, 341 & 0.2 & 40 \\
		mmWave & 123, 232 & 1 & 5 \\
		\hline
	\end{tabular}
\end{table}

\begin{figure}[!t]
	\centering
	\subfigure{\label{r1a}\includegraphics[width=.8\columnwidth]{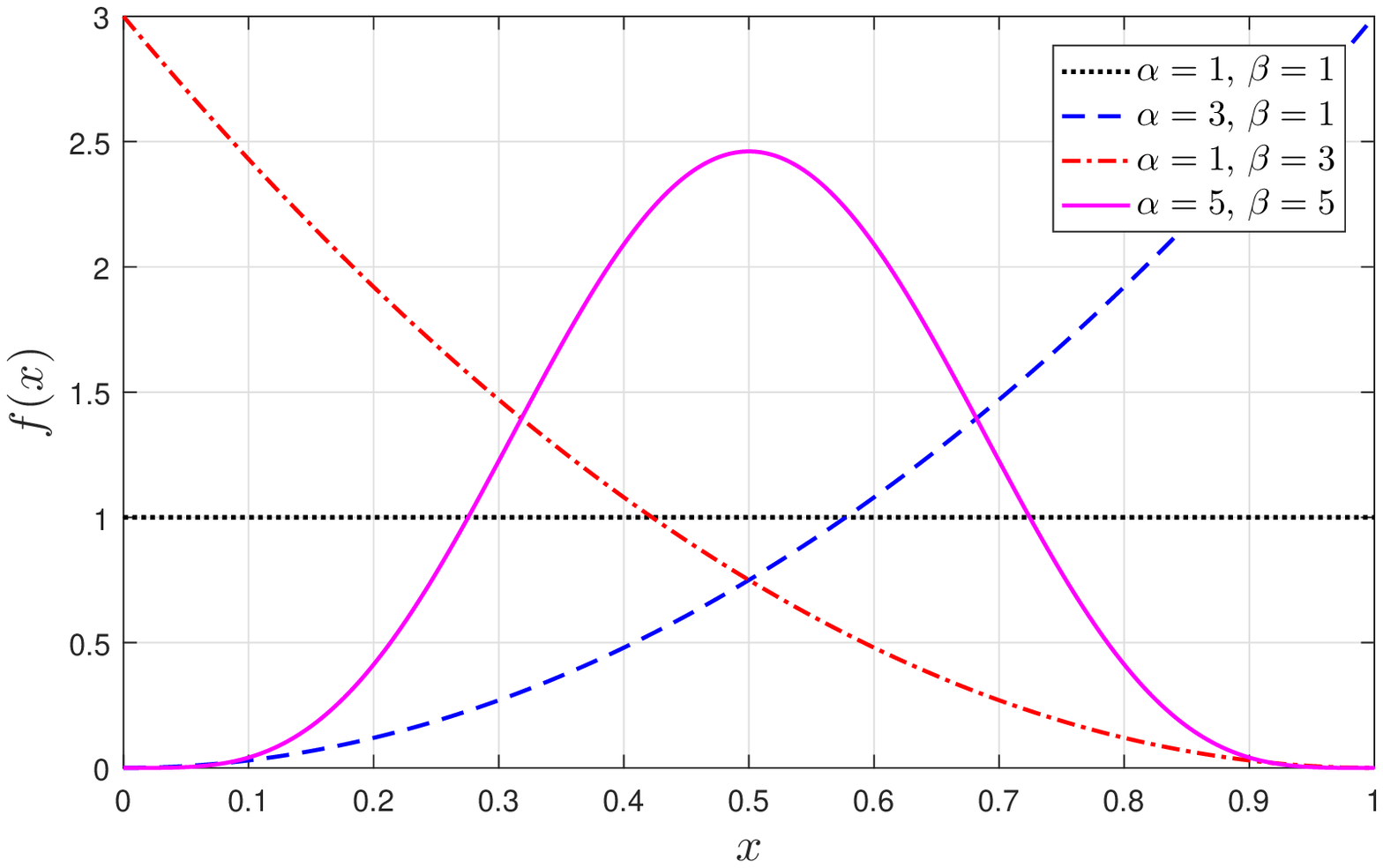}}	
	%\vspace{-4mm}
	\caption{Mean additional traffic for each link while allocating the new user load for $(\alpha,\beta)\in\{(1,3),(3,1)\}$, $m=4$ and a) $p_1$ (top), b) $p_2$ (below).}
	\label{fig_beta}
	\vspace{-2mm}
\end{figure}

We model the link mean occupations as beta-distributed random variables with probability density function (PDF)
$f(x)=\frac{\Gamma(\alpha+\beta)}{\Gamma(a)\Gamma(b)}x^{a-1}(1-x)^{b-1}$ 
since such distribution is defined in $[0,\ 1]$ and allows selecting low, heavy or even uniform loaded configurations by tuning the $(\alpha,\beta)$ parameters. $\Gamma(\cdot)$ is the gamma function. Fig.~\ref{fig_beta} shows the PDF for different pairs $(\alpha,\beta)$, and notice that ($\alpha=1$, $\beta=3$), ($\alpha=3$, $\beta=1$) model low-loaded, heavy loaded traffic configurations, respectively, while $\alpha=\beta>1$ configurations model bell-type PDFs centered at 0.5, and uniform profiles are easily obtained by setting $\alpha=\beta=1$.
\begin{figure}[!t]
	\centering
	\subfigure{\label{r1a}\includegraphics[width=.8\columnwidth]{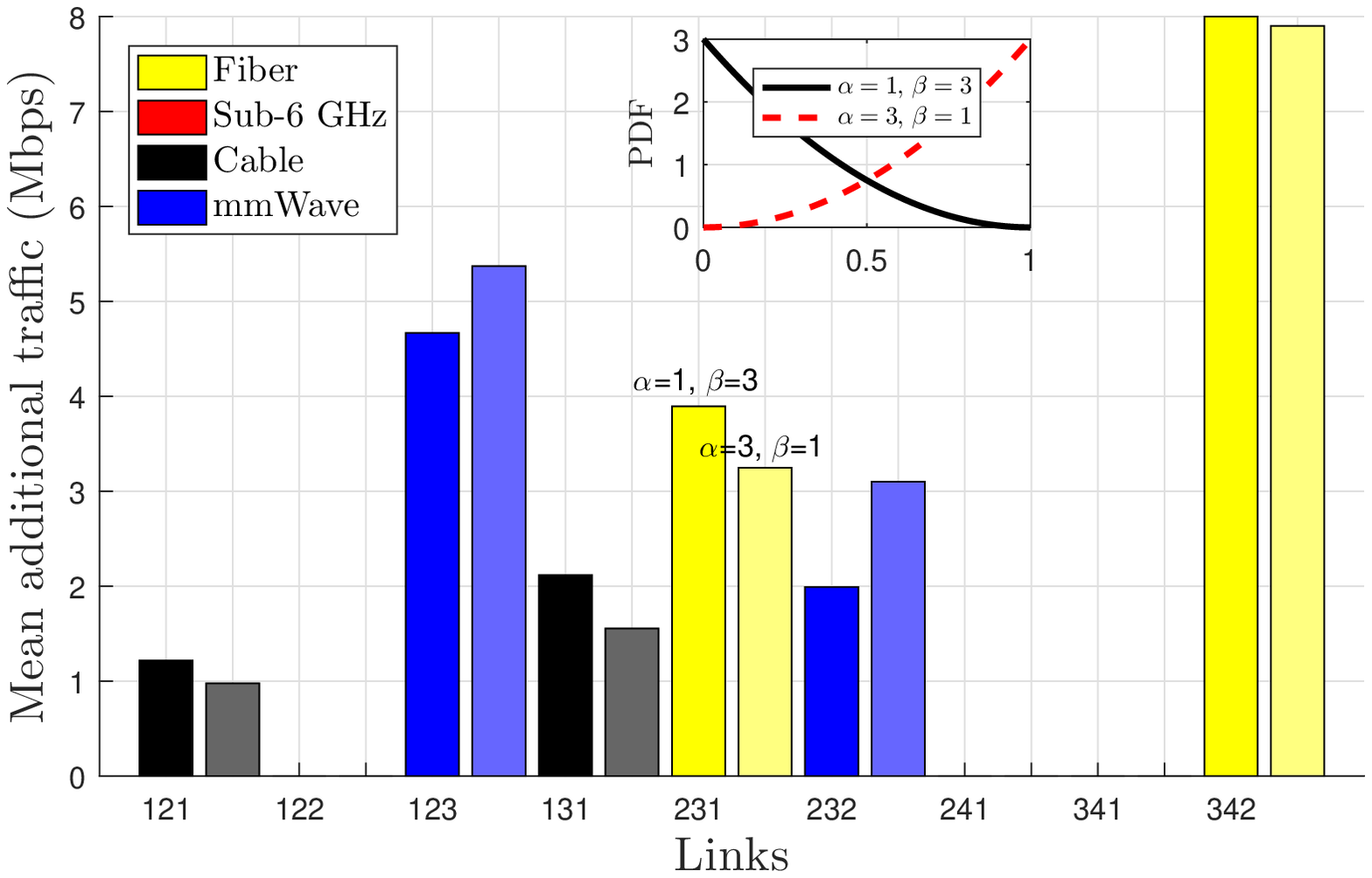}}
	%\vspace{-5mm}
	\subfigure{\label{r1b}\includegraphics[width=.8\columnwidth]{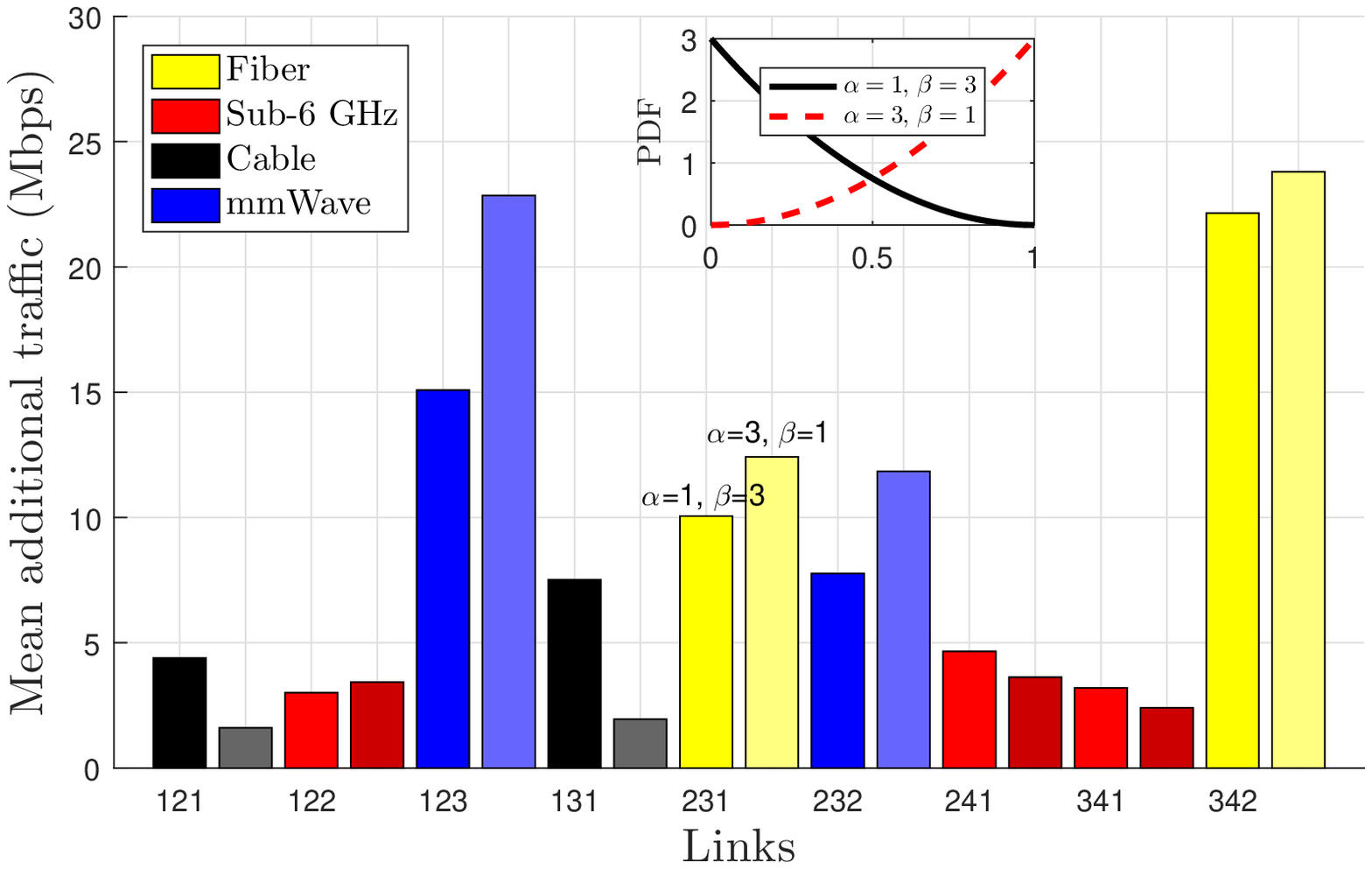}}
	\caption{Mean additional traffic for each link while allocating the new user load for $(\alpha,\beta)\in\{(1,3),(3,1)\}$, $m=4$ and a) $p_1$ (top), b) $p_2$ (below).}
	\label{fig_r1}
	\vspace{-5mm}
\end{figure}

Let a new user, e.g., a mobile device, be served by node 1 (RRH), while requiring the services of node 4 (CRAN). Fig.~\ref{fig_r1} shows how the traffic of the new user occupies each of the links, for low-loaded ($\alpha=1$, $\beta=3$) and heavy-loaded ($\alpha=3$, $\beta=1$) traffic configurations, with $m=4$ and QoS profiles: $p_1$ (Fig.~\ref{fig_r1}a) and $p_2$ (Fig.~\ref{fig_r1}b). The high mean delay of the sub $6$~GHz interfaces (in-band LTE) causes that no user with low-delay requirements could use it as a backhaul option, as shown in Fig.~\ref{fig_r1}a. The mmWave and fiber interfaces are the most used due to their simultaneous low-delay and high-throughput characteristics. Notice that all the available links are being used when the required delay is not so restrictive, as shown in Fig.~\ref{fig_r1}b. Also, the traffic dynamics of the topologies, e.g., coming from the values of $\alpha$ and $\beta$, do not change the performance significantly. Even though fiber and mmWave are the most used options, all links have a role to play to a greater or lesser extent. The chosen $m$ weights leverage both the delay and link occupations when taking the best path decisions as we discuss next.
\begin{figure}[!t]
	\centering
	\subfigure{\label{fig3a}\includegraphics[width=.8\columnwidth]{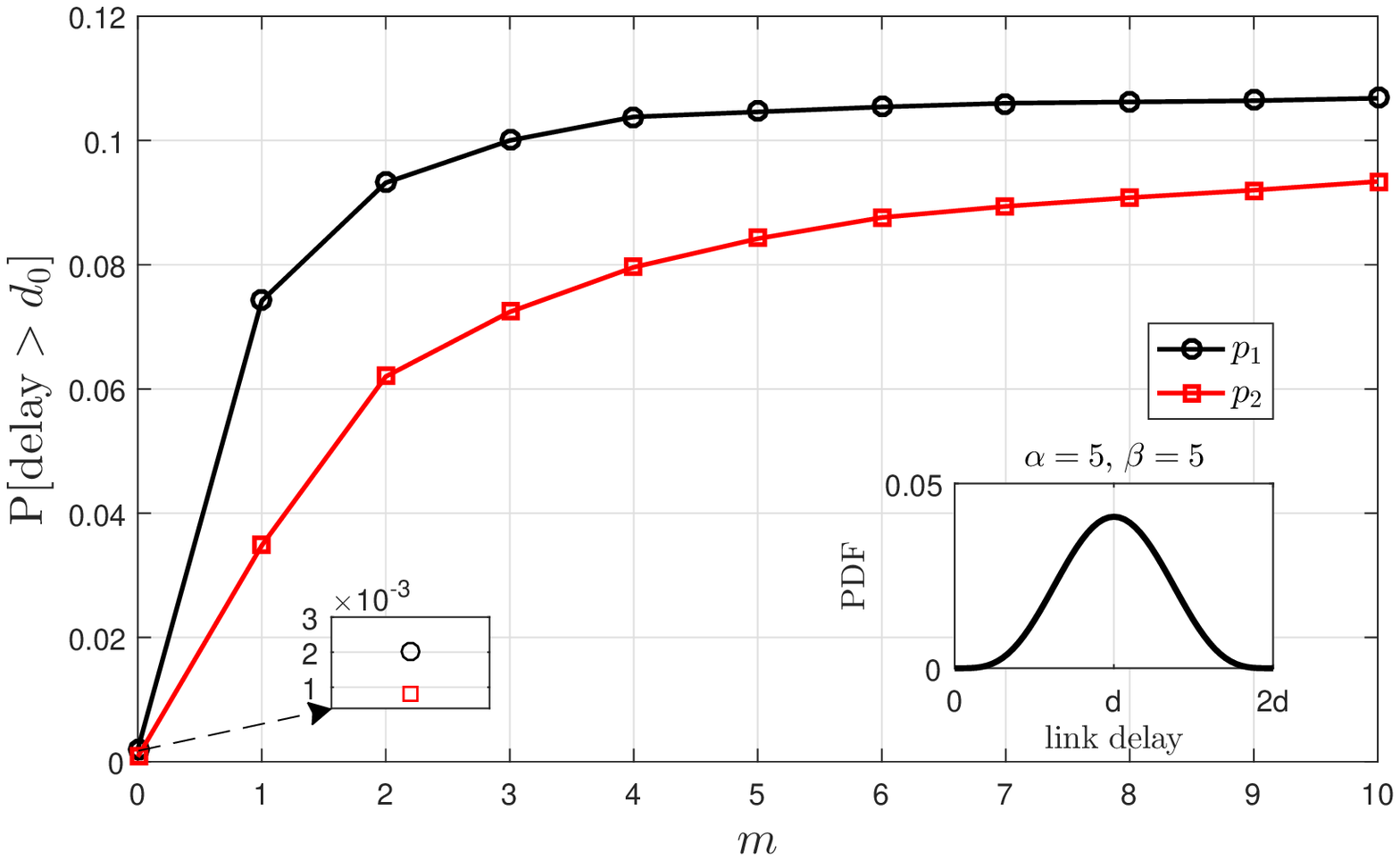}}\\
	%\vspace{-5mm}
	\ \  \subfigure{\label{fig3b}\includegraphics[width=.8\columnwidth]{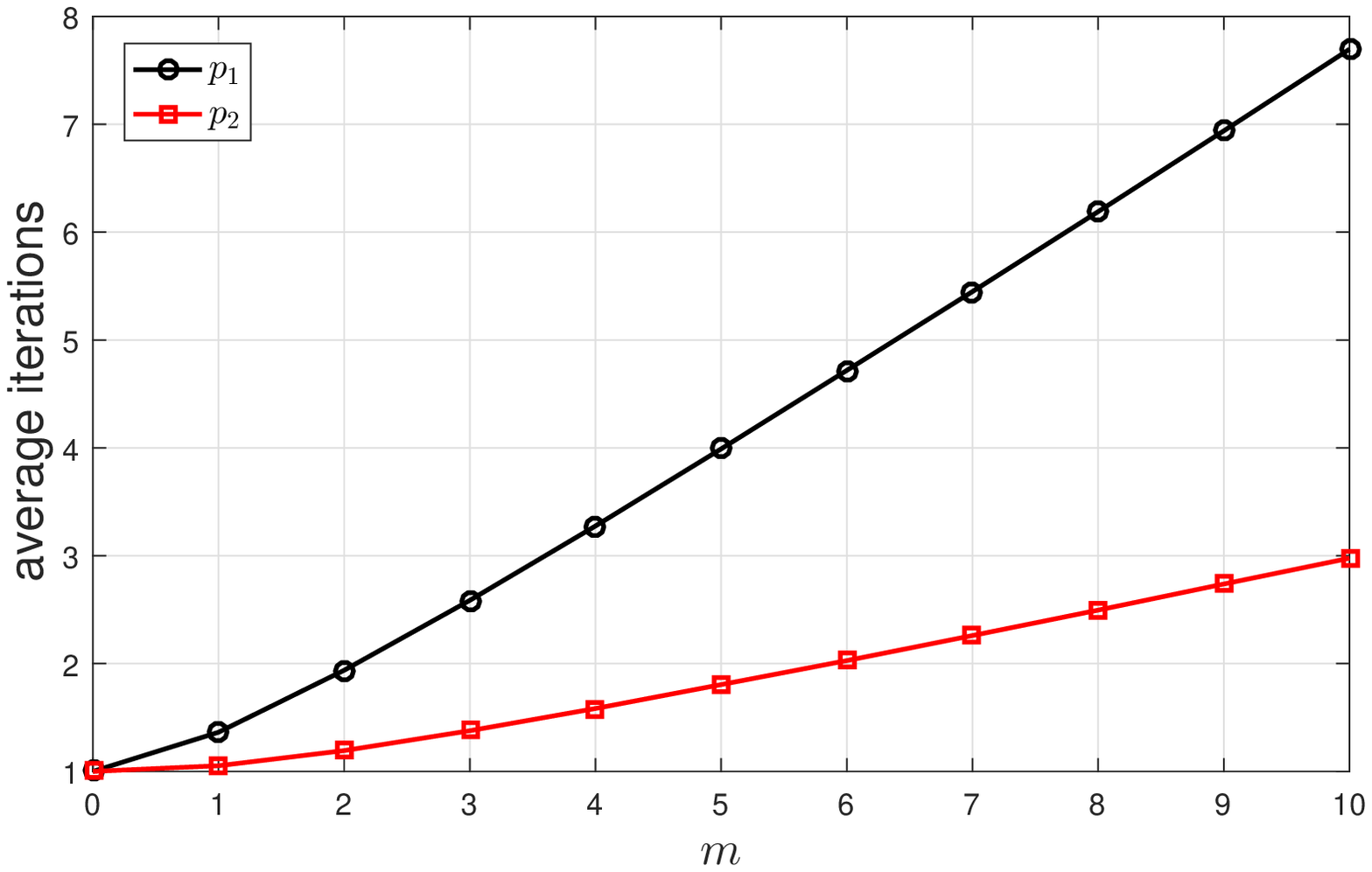}}
	\caption{Performance as a function of $m$. a) Probability of delay QoS violation for $p_1$ and $p_2$ user profiles, uniform distribution of the mean occupation of the links, and random instantaneous delays scaled by a beta distribution with $\alpha=5$, $\beta=5$, as shown in the right corner (top); b) average number of iterations when running Algorithm~\ref{alg_1}. }
	\label{fig_r3}
	\vspace{-5mm}
\end{figure}

Fig.~\ref{fig_r3} shows the impact of $m$ on the system performance.  The probability of the delay QoS violation is shown first in Fig.~\ref{fig_r3}a, where the instantaneous delay is modeled as a random variable with a bell-type PDF (scaled beta-distributed with $\alpha=\beta=5$) and mean delay specified in the rightmost column of Table~\ref{tb:delay-tput}.
Since $m=0$ leads to the path with the minimum delay, then the violation probability is the lowest as possible. However, a relatively small $m$ could take the algorithm to route the traffic through heavy occupied links, which are most likely to be saturated at the moment of its use. On the other hand, a relatively large $m$ is not advisable either, since the probability of the delay QoS violation could be severe. The more stringent the delay QoS requirement is  (either throughput or latency in this context), the grater the chances of QoS violation. The impact of $m$ also depends on the available number of interfaces on each link, since the larger they are, the higher the probability that a small change in $m$ alters the selected interface. On the other hand, $m$ also impacts strongly on the velocity of convergence of Algorithm~\ref{alg_1}, hence on the required execution time. Notice that, Algorithm~\ref{alg_1} always converges to find the required path for the coming traffic; however when $m$ increases the chances of finding the path for the required delay constraint, decrease, and more iterations are required by decreasing the value of $m$. This situation is illustrated in Fig.~\ref{fig_r3}b. Of course, the more stringent the QoS constraints, e.g., $p_1$ profile, more iterations are required on average. Observe that the number of iterations increase almost linearly with $m$.

\begin{figure}[!t]
	\centering
	\includegraphics[width=.8\columnwidth]{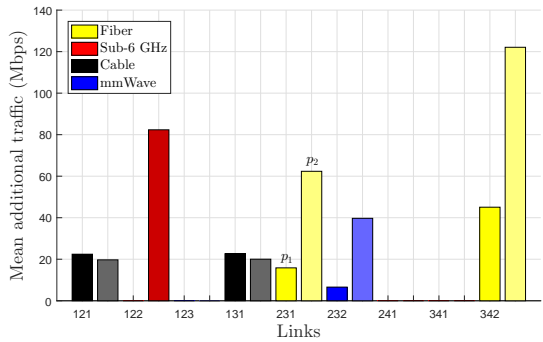}
	\caption{Mean additional traffic for each link while reallocating the load of the mmWave link $123$ in fault, for uniform distribution of the mean occupations and $m=4$.}
	\label{fig_r2}
	\vspace{-5mm}
\end{figure}

The convenience of a hybrid backhaul with multi-interfaces is also shown in Fig.~\ref{fig_r2} in case of links failures. In this particular case the mmWave link $123$ is in fault and we show how its traffic redistributes through the network. We assume a uniform distribution of the mean occupation of the links, $\alpha=1,\ \beta=1$, and that all users being served have two profiles: low-delay or high-throughput, respectively $p_1$ and $p_2$. The first case denies the choice of the sub $6$~GHz interface $122$, thus all the $p_1$ traffic must go through the cable interfaces $121$ and $131$, which are insufficient to meet the capacity demand. However, when users requiring load reallocation have $p_2$ QoS requirements, the cable interfaces along with sub $6$~GHz are able to cope with the capacity demand, carrying the traffic to the BBU and the other RRH to be then routed to the C-RAN through the mmWave and fiber interfaces.

\vspace{-2mm}
\section{Conclusion}\label{sect:remarks}
Future wireless networks will become more dense and heterogeneous, while it is necessary to guarantee connectivity and capacity in a cost-effective and sustainable way to all kinds of applications. A robust backhaul network is therefore mandatory, and a hybrid wired/wireless deployment is a potential optimum solution. The different backhaul technologies may complement each other allowing the different QoS requirements to be met at the same time. Moreover, as our results show, this can be achieved even in the advent of link failures. However, designing such dynamic, flexible and heterogeneous network brings great challenges. 
The design and coexistence of multiple-RATs and wired and wireless solutions is still an open problem and one promising alternative to tackle such diversity issue is via distributed, self-optimized networks where user association and backhauling are dynamically integrated~\cite{JaberSONACCESS16}. Another challenge is the lack of an efficient analytical framework for such heterogeneous UDNs, which enables evaluation and assessment of different strategies for backhauling. 
Despite the recent advances and architectural changes in wireless networks design, one of the greatest challenges in 5G, which is foreseen as well for future deployments, is the strict low latency requirements for broadband backhaul, specifically C-RAN applications. Up to now, only fiber and mmWave are capable of delivering few hundreds of micro seconds delay. As previously discussed, direct fiber is not pervasively available, and there is even lack of infrastructure for current networks \cite{ART:Allen-Vodafone14a}, besides the high deployment and operational costs. On the other hand, mmWave is an emerging technology that faces challenges in terms of propagation, which raises the need and motivation for further research. 
As future work, we aim to investigate different topology, RATs and incorporate other network parameters (e.g. maximum delay, reliability, minimal rate) into the algorithm, and look at distributed implementations as well. 
\vspace{-1mm}
\section*{Acknowledgments}
This work is partially supported by Academy of Finland (Grant n.303532 and n.307492), by the Finnish Funding Agency for Technology and Innovation (Tekes), Bittium Wireless, Keysight Technologies Finland, Kyynel, MediaTek Wireless, Nokia Solutions and Networks and Capes/CNPq (Brazil).
\vspace{-3mm}
\bibliographystyle{IEEEtran}
\bibliography{IEEEabrv,refs}

\end{document}